*Preprint. Subject to Revision. Comments Welcome*

# Does Artificial Intelligence Advance Science?

Liangping Ding [a], Cornelia Lawson [a,] and Philip Shapira [a,b]

liangping.ding@manchester.ac.uk
**0000-0001-6832-4114**
cornelia.lawson@manchester.ac.uk
**0000-0002-1262-5142**
pshapira@manchester.ac.uk *
**0000-0003-2488-5985**

[a] Manchester Institute of Innovation Research, Alliance Manchester Business School, The University of Manchester, UK.
[b] School of Public Policy, Georgia Institute of Technology, USA.

* Corresponding author.

June 3, 2026

## Abstract

This paper examines whether and how artificial intelligence (AI) advances scientific creativity. Drawing on scientific publications, the primary output of researchers, we analyze over one million publications from OpenAlex to investigate the relationship between AI adoption and multiple dimensions of scientific creativity, including novelty (recombinant novelty and object novelty) and impact (3-year short-run citation impact and 10-year long-run citation impact). We find that AI publications are significantly more likely to achieve top-decile creativity relative to non-AI publications, with 5.5 to 10.2 percentage point higher likelihood to rank in the top creativity decile. Critically, we uncover substantial heterogeneity across AI research modes. Tool-oriented AI research, which applies existing AI models to domain tasks, is associated with the largest gains in recombinant-based creativity, while Adaptation-oriented AI research, modifying AI models for domain-specific problems, is associated with relatively higher object-based creativity. These findings reveal that AI does not advance science through a single mechanism but through structurally distinct creative pathways that depend on how AI is incorporated into the research process. Our results contribute to ongoing debates about AI's role in science and carry direct implications for research evaluation and science policy, highlighting the need for assessment frameworks that can distinguish between recombinant and conceptual forms of creativity and that recognize how different modes of AI adoption produce fundamentally different types of scientific contribution.

# 1 INTRODUCTION

Artificial intelligence (AI) is commonly defined as the ability of machines to perform cognitive functions traditionally associated with human intelligence, including perception, reasoning, learning, problem solving, decision-making, and, in some contexts, creative tasks (Rai et al., 2019). AI is rapidly diffusing across scientific research (Duede et al., 2024), reshaping how knowledge is produced, evaluated, and rewarded. Machine learning and deep learning techniques are now routinely used for data analysis, prediction, and pattern recognition (Bryant et al., 2022). Recent advances in large language models (LLMs), particularly generative AI, have further expanded AI's role in science (L. Ding et al., 2025a), including into literature review, coding assistance, manuscript preparation, and hypothesis generation and experimental design (Hoffmann et al., 2024; Novy-Marx & Velikov, 2025). These developments have prompted widespread interest among scholars, funding agencies, and policymakers in the implications of AI for the organization and direction of scientific research (Agrawal et al., 2019; Cockburn et al., 2018; OECD, 2023).

Much of the existing literature emphasizes the productivity-enhancing potential of AI in science (Hao et al., 2026). AI tools can accelerate research workflows, reduce the cost of experimentation, and expand the space of feasible inquiry, particularly in data-intensive domains such as chemistry, materials science, and biomedicine (Stokes et al., 2020). At the same time, growing concerns have emerged regarding the broader consequences of AI adoption for scientific quality, originality, and research evaluation (Shapira et al., 2025). These concerns have become especially salient with the diffusion of large language models, which enable the rapid generation of text, code, and ideas at unprecedented scale (Kusumegi et al., 2025).

Against this backdrop, a fundamental question remains insufficiently understood: how is the use of artificial intelligence in scientific research associated with advances in the creativity of scientific outputs? In science, as we discuss subsequently, novelty refers to producing something new or different, while creativity is about generating ideas that are not only new but also meaningful, useful, and capable of advancing understanding (Heinze et al., 2009). Given that science progresses not by the volume of output but by the creativity of ideas (Foster et al., 2015), it is essential to understand how emerging technologies such as artificial intelligence interact with these creative dimensions of research. As AI-related research has emerged as a strategic priority (Agrawal et al., 2019; Cockburn et al., 2018), it has attracted substantial public and private investment. Recent evidence from NIH funding decisions suggests that proposals assisted by AI systems are more likely to receive funding, yet tend to be less distinctive relative to prior research, raising concerns that AI tools may shift scientific attention toward more incremental rather than novel ideas (Kozlov, 2026; Qian et al., 2026). From a policy perspective, establishing how AI relates to observable creative outcomes is therefore critical for assessing whether current public investments and support strategies are aligned with long-term scientific progress.

This paper makes three contributions to the literature. First, we clarify the distinction between novelty and creativity in the context of AI-assisted research and articulate why this distinction matters for science policy and research evaluation. Second, we develop a scalable empirical strategy to measure creativity using large-scale bibliometric data, while explicitly addressing data quality and classification challenges inherent in open data sources. Third, we explore the relationships between AI, novelty, and creativity in science to probe whether AI's use may influence the composition of scientific novelty without necessarily delivering creativity, differentiating between different measures of novelty/creativity. We argue that AI-related research is highly heterogeneous, and its effects on scientific creativity depend critically on how AI is positioned within the research process. Rather than treating AI as a uniform technological input, it is more appropriately conceptualized as a layered scientific infrastructure whose role varies across research contexts. Following prior work (Ding et al., 2025b), we distinguish these analytically distinct modes of AI-related research: Foundational, Adaptation, and Tool-oriented modes. These modes differ in the degree to which AI knowledge is integrated, transformed, and therefore imply different pathways to creative outcomes.

Using large-scale empirical evidence of over 1 million scientific publications this paper shows that AI-related research is associated with higher levels of creativity. Moreover, when distinguishing between different modes of AI use, we find that research that uses AI as a tool is associated with higher recombinant novelty and creativity, but shows lower association with object-based novelty, i.e., the introduction of new phrases. Instead, research that adapts AI for use in research shows the highest impact, object-based novelty and associated creativity. These results suggest that current AI primarily functions as a technology that expands combinatorial search within existing knowledge structures when used as a tool in research, rather than one that generates fundamentally new conceptual objects. However, we show that this combinatorial advantage is not uniform across research contexts. In particular, the association of AI with object-based creativity is greater when AI is adapted and developed to meet specific needs of research inquiry.

# 2 RELATED RESEARCH

## 2.1 Creativity, novelty and impact in science

We conceptualize scientific creativity as the production of knowledge that is both novel and useful (Amabile, 2018; Runco & Jaeger, 2012). In this framework, novelty refers to the extent to which a scientific contribution departs from existing knowledge. Usefulness captures the degree to which that contribution is valued within the scientific community, for example through its recognition, adoption, or subsequent use by others. Thus, while knowledge and ideas that are new and original are required for creativity, they are not sufficient. They need to be effective and valuable to the scientific community or beyond (Runco & Jaeger, 2012).

Building on this conceptual distinction, the literature has proposed empirical indicators that allow us to unpack novelty and usefulness, and thus to measure creativity (Leahey et al., 2023; Lee et al., 2015). However, novelty in scientific research is not a unified concept and there are different perspectives on its empirical measurement. An early definition by Stein (1953) suggests that creativity "arises from a reintegration of already existing materials or knowledge", thus aligning with a recombinant view of novelty where novelty is rarely conceptualized as the creation of ideas ex nihilo, but rather as the recombination of existing knowledge elements in atypical or unconventional ways (Fleming, 2001; Schumpeter, 1983; Uzzi et al., 2013). However, Stein (1953) continues that "when it is completed it contains elements that are new" (p. 311), where the extent of deviation from prior knowledge constitutes the degree of novelty. Novelty in science can encompass new theories, new ideas, new empirical findings, and new methods or instruments (Heinze et al., 2009; Leahey et al., 2023).

The second part of the definition of creativity is equally difficult to observe. The usefulness of new knowledge, ideas or tools in science relates to their effectiveness, communication and diffusion, including whether knowledge is relevant, ideas tenable or tools valuable (Runco & Jaeger, 2012). The acceptance and usefulness of such contributions are commonly reflected in measures of scientific impact (Sternberg & Lubart, 1999). While some advances in scientific creativity may be recognized relatively soon after they are communicated, in other cases scientific creativity becomes evident only through later recognition (Fleming, 2001). Where scientific community recognition occurs after a delay (Wang et al., 2017), the literature has termed such instances as "sleeping beauties" (Ke et al., 2015; Van Raan, 2004). Moreover, while usefulness can be associated with high novelty, the two are not equivalent, and highly novel research strategies frequently entail substantial risk and uncertain usefulness (Foster et al., 2015).

## 2.2 AI and scientific knowledge

Artificial intelligence has the potential to reshape the processes through which scientific knowledge is generated, recombined, and evaluated, thereby influencing creativity in research. A growing body of empirical evidence suggests that AI's effects on knowledge work are uneven and contingent on the alignment between task requirements and technological capabilities. Dell'Acqua et al. (2023), for example, show that AI introduces a "jagged technological frontier," whereby tasks that fall within AI's competence exhibit substantial improvements in productivity and quality, while tasks outside this frontier remain difficult to complete successfully. This finding underscores that AI does not uniformly enhance performance across all activities but instead differentially augments specific components of complex work. This extends to scientific research, which is composed of multiple interdependent tasks, including problem formulation, conceptual development, data analysis, and communication, that differ in uncertainty, decomposability, and coordination requirements (Fleming & Sorenson, 2004; Puranam et al., 2014). AI may complement some of these tasks while substituting for others, with uncertain implications for scientific creativity.

Evidence from the science of science literature further suggests that AI adoption can substantially reshape scientific output and influence. Hao et al. (2026) suggest that scientists who engage in AI-augmented research publish more papers, receive more citations, and assume leadership roles earlier in their careers compared to those who do not adopt AI tools. These findings are consistent with theories that tools lowering search and production costs can increase individual productivity and accelerate knowledge accumulation (Jones, 2009). However, in the same study, Hao et al. also find that AI adoption reduces the collective diversity of scientific topics. This pattern points to a potential tension between individual-level gains and system-level creativity: while AI may enable researchers to produce more influential work, it may simultaneously concentrate attention on narrower problem domains and reduce exploratory diversity, both of which are critical drivers of breakthrough innovation (Fleming, 2001).

Organization and innovation research provides additional insight into the mechanisms underlying these dynamics. Technologies that codify expertise and reduce information frictions can broaden participation in knowledge production and alter collaboration structures (Ding et al., 2010). AI systems may function as cognitive intermediaries that substitute for certain forms of interpersonal coordination while complementing others (Hoffmann et al., 2024), potentially reshaping how teams generate and integrate ideas. Reduced reliance on peer interaction could lower opportunities for serendipitous recombination – long recognized as an important source of creativity – while increased individual autonomy may accelerate iterative experimentation. In this sense, AI may simultaneously enable more efficient local search while constraining distant search across the knowledge space (March, 1991).

Taken together, existing evidence is mixed, suggesting that AI can both enhance and constrain the creativity of scientific knowledge production through multiple mechanisms: by expanding individual capabilities, altering search processes, reshaping collaboration patterns, and influencing the distribution of attention across research domains. These competing forces imply that the net effect of AI on scientific creativity is theoretically ambiguous and clearly calls for additional empirical investigation.

### 2.3 Modes of AI use and creativity

In disentangling AI's impacts on science, it is important to distinguish *how* AI is being used. AI is not uniformly applied in science and we posit that its effects on scientific creativity depend critically on how it is positioned within the research process. Returning to the definition of creativity, we can consider AI that contributes towards new methods or instruments, new results or empirical phenomena, and new theories and ideas as contributing towards creativity (Heinze et al., 2009; Leahey et al., 2023). The most *foundational* form of AI in this context is the provision and development of new AI methodologies, which are able to support the empirical testing of theoretical problems (by others) (Heinze et al., 2009). Here, creativity arises from the introduction of new problem framings, technical vocabularies, and representational structures that are valuable for addressing complex empirical and analytical problems.

However, in most domains of science, AI primarily functions as a methodological *toolkit* that can be combined with domain-specific knowledge to address complex empirical and analytical problems (OECD, 2023). AI technologies have the potential to substantially expand the set of analytical techniques available to researchers and lower the cost of integrating computational methods with domain-specific problems, thereby increasing opportunities for creativity (Cockburn et al., 2018; Rai et al., 2019). This corresponds to Krenn et al. (2022)'s first dimension of AI assisted scientific research, which emphasizes AI's role as a *tool* that enhances measurement and analysis to reveal new results and insights. While this may lead to new and surprising results, AI as a *tool* may primarily automate existing processes or substitute experimental means, without resulting in the uncovering or creation of new scientific entities or concepts.

At the interface, scientists may *adapt* and develop AI to apply to domain-specific problems 6/3/2026 6:13:00 PM with the goal of uncovering hidden dimensions in data. Such a more targeted use of AI may increase its potential to act as what Krenn et al. (2022) termed a 'resource of inspiration', to inform new ideas and theories. Creativity thus arises from AI's modular and *adaptable* nature, that facilitates the recombination of new computational approaches and domain-specific research questions (Agrawal et al., 2019).

By distinguishing among these three modes – foundational, tool, and adaptation – we suggest that AI's relationship with scientific creativity is unlikely to be homogeneous. Instead, creative outcomes are contingent on how AI is integrated into the knowledge production process, influencing the mechanisms through which novelty and impact emerge.

# 3. DATA SET AND VARIABLE CONSTRUCTION

## 3.1 Data

To investigate how AI adoption relates to the scientific creativity of research output, we use scientific publications — researchers' primary research products — as our unit of analysis. Scientific publications represent the creative output of researchers, documenting the research process in sufficient detail to support reproducibility, and AI technology employed as a substantive research methodology is typically disclosed within the publication. However, the mere mention of AI in a paper's text cannot be directly equated with AI adoption: papers such as literature reviews or perspective pieces discuss AI as a subject of inquiry rather than employing it as a methodological tool (Ding et al., 2025a). Furthermore, the specific mode of AI use may shape scientific creativity, for example, whether an off-the-shelf AI system is applied to a problem versus whether a novel AI model architecture is developed and adapted to the task at hand.

To address these distinctions, this study combines traditional bibliometric methods with the AI modes framework (Ding et al., 2025b) to classify publications into AI-adoption publications and those that do not adopt AI (hereafter *non-AI publications*). We further classify AI-adoption publications into three AI modes: Adaptation, Tool, and Foundational.

The classification proceeds in two stages. In the first stage, we assess the AI relevance of each publication by combining bibliometric keyword search with a GPT-SciBERT pipeline. This approach not only identifies AI-related publications based on a curated AI keyword set but also filters out false positives by leveraging semantic embeddings of titles and abstracts. In the second stage, we apply the AI modes framework developed by Ding et al. (2025b). That work categorizes AI-related publications into four modes: Adaptation, Tool, Foundational, and Discussion, the latter comprising papers that refer to AI without employing it in the research process. In the present study, we focus on the actual use of AI in scientific research. Accordingly, publications classified as Adaptation, Tool, or Foundational are grouped as AI-adoption publications, whereas those classified as Discussion are not included in the analysis.

The primary data source for this study is OpenAlex (Priem et al., 2022). OpenAlex contains over 450 million bibliometric records drawn from PubMed, CrossRef, arXiv, and other sources, providing comprehensive metadata on scientific publications, including titles, abstracts, DOIs, publication venues, publication years, and field classifications. Compared to Web of Science and Scopus, OpenAlex offers considerably better coverage of scholarly publications, especially conference proceedings. This makes it particularly well suited for our study for two reasons. First, conference proceedings are a major communication channel in AI research given the field's rapid pace of development; OpenAlex's coverage of this venue type helps capture a more representative picture of the AI landscape and mitigates selection bias. Second, OpenAlex is a product of the Open Science movement and benefits from an active research community that continuously enriches the database by linking it to additional publication-level indicators. The combination of broad coverage and active community development makes OpenAlex the most appropriate foundation for our analytical dataset. The OpenAlex snapshot used in this study was retrieved on May 30, 2025.

In the first stage of classification, we construct the AI-relevant publication sample following Ding et al. (2025b), restricting the analysis to peer-reviewed journal articles and conference proceedings.[1] Candidate AI-relevant publications are first identified by applying the AI keyword set from Ding et al. (2025b) to the concatenated titles and abstracts of all publications in OpenAlex. We then apply a series of preprocessing filters to refine the candidate pool: we restrict to English-language publications with valid DOIs; exclude retracted papers and non-research outputs such as editorials and commentaries; remove records with missing titles or abstracts; and eliminate duplicates based on DOI, title, and abstract. The sample is further restricted to publications from 2004 to 2024, a period that captures the contemporary development and diffusion of AI technologies across scientific disciplines. To reduce false positives introduced by keyword matching, we apply the two-stage GPT-SciBERT filtering pipeline from Ding et al. (2025b), which generates high-dimensional semantic embeddings of titles and abstracts and thereby captures topical relevance beyond keyword boundaries. In the second stage, the stage one papers are classified into the AI modes of Adaptation, Tool, and Foundational, with Discussion-mode publications dropped. The publications classified under the Adaptation, Tool, and Foundational modes form the AI-adoption group.

[1] We exclude preprints and other non-peer-reviewed outputs for two reasons. First, citation and reference information for preprints is often incomplete in bibliometric databases, which could lead to inaccurate creativity scores. Second, since AI-related research is disproportionately represented in preprint form, retaining them would introduce systematic bias and likely underestimate AI's true impact.

Given that the OpenAlex database in our study contains approximately 221 million non-AI records, constructing a comparable non-AI sample poses substantial computational challenges. We address this through stratified random sampling designed to yield a non-AI comparison group that is representative of the broader disciplinary landscape of non-AI research. Specifically, we randomly sample 2% of non-AI-relevant publications with complete title and abstract information, stratifying by each paper's primary scientific field using the 26 OpenAlex field classifications to preserve the disciplinary distribution of research output. This procedure yields an initial non-AI sample of 2,241,477 publications. We then apply the same preprocessing filters as in the first stage of AI relevance classification, which includes restrictions on language, document type, publication year, DOI availability, and deduplication.

To accommodate missing values in our main dependent, we further restrict the sample to observations with non-missing 10-year citation impact and recombinant novelty scores.[2] After preprocessing, the final analytical dataset comprises 1,127,716 publications, of which 697,551 are AI-adoption publications and 430,165 are non-AI publications. The illustration of the analytical dataset construction pipeline is shown in Figure 1.

[2] Missing values in 10-year citations arise from a discrepancy between SciSciNet and the OpenAlex version used in our study, with some publications not included in the former. Missing values in recombinant novelty come from missing reference information and thus the inability to calculate such measure. The sample attrition from these restrictions is however minimal.

### 3.2 Dependent Variables

In line with the extant literature, we measure creativity as the combination of novelty and impact. Novelty captures the extent to which a scientific contribution departs from existing knowledge, while impact reflects the degree to which novel ideas are adopted, recognized, or built upon by subsequent research. Building on the classification framework proposed by Shibayama et al. (2025), we distinguish novelty measures by whether they adopt recombination-based or object-based approaches. Recombination-based measures define novelty as atypical combinations of existing knowledge elements such as references, journals, or disciplinary categories (Lee et al., 2015; Wang et al., 2017). Object-based measures capture novelty through the introduction of new entities, keywords, or concepts that were previously absent from the scientific record (Arts et al., 2025).

While novelty indicators are typically measured ex-ante at the time of publication using observable features such as reference combinations or textual content, impact needs to be considered ex-post, inferred retrospectively from downstream uptake, most notably citations (Bornmann & Daniel, 2008; Garfield, 1972). Citations capture the extent to which subsequent research acknowledges, adopts, or builds upon a focal contribution, making them a widely used – albeit imperfect – measure of scientific value (Waltman, 2016).

#### *3.2.1 Novelty*

Following Lee et al. (2015), we measure recombinant novelty based on the atypicality of publication outlets in a paper's reference list. A paper is considered highly novel if it combines references from outlets that are rarely cited together within the scientific literature. Compared to the Uzzi et al. (2013) measure, which uses pairwise journal co-occurrence z-scores and produces highly skewed distributions, the Lee et al. (2015) measure is less computationally intensive and produces a distribution closer to normality. Novelty scores for the focal publication are computed using the novelpy package (Pelletier & Wirtz, 2023), with reference data drawn from all publications indexed in OpenAlex across all publication outlets. Following prior studies (Bianchini et al., 2026; Wang et al., 2017), we construct a binary indicator equal to one if the publication ranks in the top 10 percent of the novelty distribution within its field. To address sparsity issues in some OpenAlex fields, we use the OECD field classification when defining field-specific novelty thresholds. We map the 26 OpenAlex fields to the OECD Fields of Science and Technology (FOS) classification,[3] aggregating them into seven broad scientific domains, including Natural Sciences, Engineering, Medical Sciences, Agricultural Sciences, Social Sciences, Computer science and Humanities.

We measure object-based novelty using the measure provided by Arts et al. (2025) based on noun phrases extracted from the title and abstract of each publication. Object novelty is operationalized as a binary variable equal to one if the publication introduces at least one noun phrase that (i) appears for the first time in the scientific record, and (ii) is subsequently reused in at least one other publication. A percentile cutoff is not appropriate here because the majority of publications introduce no new reused phrases, producing a distribution with excessive mass at zero that makes within-decile distinctions uninformative.

#### *3.2.2 Impact*

We measure scientific citation impact using forward citation counts within fixed windows of three years and ten years after publication drawn from the SciSciNet data platform, which provides pre-calculated fixed window citation impact.[4] Since citation practices vary across scientific fields and over time, we further adopt the Mean Normalized Citation Score (MNCS) to enable meaningful cross-field comparisons (Waltman & van Eck, 2013). The MNCS normalizes a publication's citation count by the average number of citations received by all publications published in the same field and year, so that a score of 1 indicates citation impact at the world average, a score above 1 indicates above-average impact, and a score below 1 indicates below-average impact. Furthermore, we identify whether each publication ranks in the top 10 percent of citation counts within its OECD field and publication year. Specifically, we employ two binary indicators corresponding to the three-year and ten-year citation windows, respectively, enabling examination of both short-run recognition and long-run influence. By focusing on the top 10 percent threshold, we complement the MNCS measure with a robust, rank-based indicator that is less sensitive to extreme outliers and provides a clearer signal of whether a publication achieves elite scientific recognition within its field. Note that publications from 2022 onwards are excluded from three-year citation analyses to ensure that sufficient time has elapsed for citation accumulation, however we include all years in the 10-year measures, which implies shorter observation windows for more recent publications.

[3] The mapping is based on the correspondence between OpenAlex fields and the Scopus All Science Journal Classification (ASJC) codes, following the classification scheme provided by Elsevier. https://service.elsevier.com/app/answers/detail/a_id/12007/supporthub/scopus/. Note that we didn't aggregate computer science into Natural sciences.

### *3.2.3 Creativity*

We construct aggregate measures of creativity by combining novelty and citation impact. Rather than requiring simultaneous top-decile membership on both dimensions which would mechanically produce very small creative sets given the low base rate of each component, we adopt a percentile-rank composite approach to arrive at the percentile rank creativity (PCRE). For each publication, the novelty score and the citation count are separately ranked into percentiles from 1 to 100 within the full estimation sample, rather than within the full population as with the underlying novelty and citation impact measures. We rank within the estimation sample because creativity is defined here as a relative concept, which captures publications that are simultaneously novel and impactful relative to the set of publications under study rather than as an absolute population-level benchmark. The two percentile ranks are then averaged to form a composite creativity score (scaled 0 to 1), and a publication is classified as highly creative if its composite score falls in the top 10 percent of the distribution. We construct four creativity indicators corresponding to two novelty types and two citation horizons: Recombinant Creativity (3-year), Recombinant Creativity (10-year), Object Creativity (3-year), and Object Creativity (10-year).

[4] We utilize the SciSciNet-v2 version, which was constructed using an OpenAlex version proximate to the one we employ. As a result, sample attrition from merging with SciSciNet is minimal and unlikely to introduce systematic bias.

### 3.3 Independent Variables

Our key independent variable is *AI adoption*, which captures whether a publication substantively engages with AI methods as part of its scientific contribution. AI-related publications are classified into *Adaptation*, *Foundational*, and *Tool* modes (see Section 3.1). This allows us to distinguish the overall association between AI adoption and scientific creativity from the heterogeneous effects associated with different epistemic roles of AI.

### 3.4 Control Variables

To account for potential confounding factors that may influence the relationship between AI adoption and scientific novelty, impact, and creativity, we include a set of control variables commonly used in the literature (Arts et al., 2025; Lee et al., 2015; Wang et al., 2017). For all models that estimate the effect on creativity, recombinant-based novelty and citation impact, we control for reference count, as these are mechanically linked to the size of the reference list. For object novelty/creativity specifications, we control for the total noun phrase count in the title and abstract to account for variation in textual richness and abstract length. In addition, we account for collaboration characteristics by including the number of authors, the number of distinct institutional affiliations, and the number of distinct countries associated with each publication. All these control variables are logged to normalize the highly skewed distributions. We also include a binary indicator for publication type (journal versus conference proceeding), as citation norms and novelty distributions differ substantially across these publication types.

### 3.5 Econometric Model

To examine the relationship between AI adoption and scientific novelty, impact, and creativity, we estimate the following baseline specification:

$$Y_i = \alpha + \beta\, AI_i + \gamma\, X_i + \alpha_t + \alpha_i + \alpha_k + \varepsilon_i$$

where $Y_i$ denotes the binary outcome of publication $i$ across the set of novelty, impact, and creativity measures defined in Section 3.2.[5] Given the binary nature of all outcome variables, we estimate probit models and report average marginal effects (AMEs). The variable $AI_i$ is the key explanatory variable, equal to one if publication $i$ adopts AI methods and zero otherwise; in mode-level specifications, $AI_i$ is replaced by a set of mutually exclusive mode dummies (Adaptation, Foundational, Tool), with non-AI publications as the reference category. The coefficient $\beta$ captures the association between AI adoption and the outcome. The vector $X_i$ contains the control variables. All regressions include publication year fixed effects ($\alpha_i$), OECD field fixed effects ($\alpha_i$), which flexibly control for unobserved heterogeneity across scientific fields and over time, such as field-specific publication norms, citation practices. Heteroskedasticity-robust standard errors are reported throughout. Detailed definitions of all variables are provided in Table A1 in the Appendix.

### 3.6 Summary Statistics

Table 1 presents summary statistics for the recombinant novelty/10-year citation impact estimation sample, which is the most complete sample in our setting. The main sample comprises 1,127,716 publications, of which 61.9% are classified as AI-adoption publications. Among AI-adoption publications, Adaptation-oriented research accounts for 25.3% of the full sample, Tool-oriented research for 30.9%, and Foundational research for 5.7%. These shares reflect the dominance of Adaptation and Tool-oriented AI research in the contemporary scientific literature.

[5] Specifically, we estimate models for Top 10% Recombinant Novelty (*Novelty Recomb.*), Object Novelty (*Novelty Object*), 3-year Citation Impact (*Impact 3-yr*), 10-year Citation Impact (*Impact 10-yr*), Recombinant Creativity-3year (Creativity *Recomb. 3yr*), Recombinant Creativity-10year (Creativity *Recomb. 10yr*), Object Creativity-3year (Creativity *Object 3yr*), and Object Creativity-10year (Creativity *Object 10yr*).

Top-decile outcome indicators display means consistent with their construction. The average rate of top 10% recombinant novelty is 17.2%, while top 10% object novelty stands at 21.4% within the relevant estimation sample. Citation impact indicators show similar means, with 30.4% of publications in the three-year impact sample ranking in the top decile, and 28.6% for ten-year impact. These rates substantially exceed 10% because the top-decile thresholds are benchmarked against the full population, and the estimation sample is a selected subset that disproportionately contains high-novelty and high-impact publications. Creativity measures, by contrast, are constructed within the estimation sample, and their rates of 9.8%–12.8% remain close to 10% by construction, with minor deviation reflecting ties in the composite score distribution. Control variables display substantial heterogeneity: the mean log reference count is 2.7, mean log number of authors is 1.3, and mean log number of affiliations is 0.65, reflecting variation in collaborative scale and reference practices across the sample. About 85% of articles were published in journals, with the remaining 15% representing conference proceedings.

------------------------------------

Insert Table 1 about here

------------------------------------

Table 2 reports mean differences between AI- and non-AI publications. AI publications are significantly more likely to be among the top 10% novel publications based on recombinant novelty, with almost 22% compared to under 10% for non-AI. This difference is observed for all modes of AI but largest for Tool, where 27% of papers are among the top 10%. Conversely, non-AI publications have higher object novelty, as they are slightly more likely to introduce a new phrase compared to AI publications. However, this difference is driven by Tool publications, which are the least likely to contain new phrases. Adaptation and Foundational papers instead are more likely to contain a new phrase compared to non-AI publications. AI publications across modes outperform non-AI publications in terms of citation impact. In terms of the association with creativity AI papers outperform non-AI papers. However, Tool papers perform lowest on creativity based on the object-creativity-10year measure. In addition, pairwise Pearson correlations among the key explanatory variables and controls are reported in the appendix Table A2. None of the reported correlations approach levels typically associated with multicollinearity concerns.

------------------------------------

Insert Table 2 about here

------------------------------------

# 4 EMPIRICAL RESULTS

## 4.1 AI Adoption and Scientific Novelty and Citation Impact

Table 3 presents average marginal effects (AMEs) from probit models for scientific novelty and citation impact. Odd-numbered columns include a binary AI indicator; even-numbered columns decompose AI-adoption research into three mode dummies, with non-AI publications as the reference category.

### *4.1.1 Overall AI Effect*

Three principal findings emerge from Columns (1), (3), (5), and (7). First, AI adoption is positively and significantly associated with recombinant novelty. The AME of the binary AI indicator in Column (1) is 0.161 ($p < 0.001$), indicating that AI-adoption publications are 16.1 percentage points more likely to rank in the top 10% recombinant novelty relative to non-AI publications within the same field and year. This is the largest of the four outcome effects and suggests that AI adoption is strongly associated with the combinatorial dimension of novelty, which includes the synthesis of diverse knowledge strands.

Second, AI adoption is significantly associated with object novelty, although the effect is smaller in magnitude. The AME in Column (3) is 0.050 ($p < 0.001$), meaning AI-adoption publications are approximately 5.0 percentage points more likely to introduce novel terminology. The comparatively modest magnitude suggests that AI-driven research produces relatively less object novelty in terms of new constructs and vocabulary, even as it recombines existing knowledge in novel ways.

Third, AI adoption is strongly associated with both short-term and long-term citation impact. In Column (5), the AME on three-year impact is 0.089 ($p < 0.001$), and in Column (7), the AME on ten-year impact is 0.096 ($p < 0.001$). The 10-year AME slightly exceeds the three-year AME, suggesting a modest intensification of AI's citation advantage over longer horizons.

#### *4.1.2 Heterogeneity by AI Mode*

Columns (2), (4), (6), and (8) reveal substantial heterogeneity across AI research modes. For recombinant novelty (Column 2), Tool-oriented publications display the largest AME (0.191, $p < 0.001$), exceeding the effect for Adaptation (0.120) and Foundational (0.103) modes. This pattern is consistent with Tool-oriented research drawing on and synthesizing knowledge from diverse computational traditions. In contrast, for object novelty (Column 4), Adaptation-oriented publications display the largest effect (0.102), while Tool-oriented exhibits the weakest effect among three AI modes (0.023). This divergence suggests that the Adaptation-mode is more likely to introduce new conceptual constructs and terminology, whereas the Tool-mode achieves novelty primarily through recombination rather than conceptual innovation. For citation impact, Adaptation-oriented publications display the highest AMEs for both three-year impact (0.120, Column 6) and ten-year impact (0.116, Column 8), suggesting that adapting AI model architecture to domain-specific tasks generates the most widely cited outputs.

Figure 2 plots the predictive margins for each AI mode across the four novelty and impact outcomes with 99% confidence intervals. All three AI modes lie substantially above the non-AI baseline across all outcomes. The visualization confirms the mode-level ordering evident in the regression estimates: Tool-mode research consistently displays the highest predicted probability for recombinant novelty, while Adaptation-mode research occupies the leading position for object novelty and both citation impact measures.

--------------------------------------------------------------------
Insert Table 3 and Figure 2 about here
--------------------------------------------------------------------

### 4.2 AI Adoption and Scientific Creativity

Table 4 presents average marginal effects (AMEs) from probit models for scientific creativity. The four creativity indicators span two novelty types (recombinant and object) and two impact horizons (three-year and ten-year).

#### *4.2.1 Overall AI Effect on Creativity*

AI adoption is positively and significantly associated with all four creativity indicators. For recombinant creativity, the AME is 0.116 ($p < 0.001$) for the three-year horizon (Column 1) and 0.103 ($p < 0.001$) for the ten-year horizon (Column 3). For object creativity, the corresponding AMEs are 0.056 (Column 5) and 0.048 (Column 7), both significant at $p < 0.001$. These magnitudes are consistent with, though somewhat smaller than, the component effects reported in Table 3. AI-adoption publications are 16.1 percentage points more likely to achieve top 10% recombinant novelty and 8.9 percentage points more likely to rank in the top 10% three-year citation impact individually. Creativity, which requires high performance on both dimensions simultaneously, is therefore rarer but still meaningfully elevated for AI-adoption research (11.6 percentage points in Table 4, column 1). The larger effects for recombinant creativity than for object creativity mirror the corresponding pattern in Table 3 and reinforce the conclusion that AI's contribution to scientific output operates most strongly through the recombinant knowledge channel. The stability though slight attenuation of effects across three-year and ten-year citation horizons suggests that AI-adoption creativity represents a durable scientific contribution rather than a citation spike driven by novelty attention.

#### *4.2.2 Heterogeneity by AI Mode*

The mode-level disaggregation in even-numbered columns reveals additional patterns. For recombinant-based creativity, Tool-oriented publications display the largest AMEs (0.126 at three years; 0.112 at ten years), consistent with the novelty results in Table 3 and confirming that research that uses AI as a tool produces outputs that are both structurally novel and highly cited. Adaptation-oriented publications follow, with AMEs of 0.103 (three years) and 0.095 (ten years), while Foundational research shows significant but smaller effects (0.084 and 0.069, respectively).

The mode ordering reverses for object creativity. Here, Adaptation-oriented publications display the largest AMEs (0.101 at three years; 0.085 at ten years), while Tool-oriented research shows more modest effects (0.033 and 0.030). This reversal is consistent with the object novelty results in Table 3. Adaptation-mode research, which modifies and reconfigures AI models to address domain-specific problems, is more likely to generate new conceptual vocabulary and domain-specific frameworks. Tool-mode research, by contrast, in applying existing AI models to domain tasks without fundamental modification, achieves novelty primarily through recombination of established knowledge structures rather than through the introduction of new concepts. Foundational AI publications again display consistently positive effects across all creativity measures but lower than adaptation.

Figure 3 plots the predictive margins by AI mode for all four creativity outcomes. The crossover pattern between recombinant and object-based creativity is visually apparent: Tool-mode research occupies the highest position for both recombinant creativity measures (three-year and ten-year), while Adaptation-mode research is associated with relatively greater object creativity. All three AI modes are substantially above the non-AI baseline, confirming that AI adoption broadly aligns with creative scientific output regardless of research mode. Taken together, the estimates in Table 4 and Figure 2 suggest that AI's contribution to scientific creativity is not monolithic: Tool-mode research operates primarily through recombinant pathways, while Adaptation-mode research exhibits comparatively stronger object creativity.

------------------------------------------------------------
Insert Table 4 and Figure 3 about here
------------------------------------------------------------

### 4.3 Robustness Checks

Five sets of robustness checks confirm the stability of the main findings. First, Appendix Table A3 replicates the creativity analysis using a factor-analysis-based creativity index (FCRE) as an alternative to the main percentile-rank composite (PCRE). For each novelty-citation pair, principal component factor analysis with varimax rotation is applied to extract a single latent creativity factor from the two inputs. Because the resulting factor scores can take negative values, continuous specifications use an Inverse Hyperbolic Sine (IHS) transformation, which approximates the natural log for large positive values while remaining defined for zero and negative scores. For binary specifications we binaries the raw factor score at its 90th percentile. Results are qualitatively and quantitatively similar to the main PCRE estimates (e.g., AMEs of 0.114 and 0.099 for recombinant creativity at three and ten years), and the mode-level ordering mirrors the baseline results, confirming that the main findings are not sensitive to the method used to combine novelty and impact.

Second, Appendix Table A4 uses the continuous PCRE score (scaled 0-1) as the dependent variable estimated by OLS, rather than the binary top-decile indicator in the main probit models. The coefficient pattern is consistent with the main results in Table 4. AI-adoption publications score higher on the continuous creativity scale across all four specifications, and the mode-level heterogeneity is preserved with Tool-mode publications exhibiting the largest positive coefficients for recombinant creativity and Adaption-mode publications exhibiting the largest positive coefficients for object creativity. This indicates that the findings are not artefacts of binarization at the 90th-percentile threshold.

Third, Appendix Table A5 excludes publications with at least one US or Chinese co-author, addressing concerns that the results may be driven by the outsized contribution of these two countries to both AI research and citation accumulation. Despite the substantially reduced sample, AI and mode-level AMEs remain positive, significant, and of comparable magnitude, indicating that the findings are not geographically concentrated.

Fourth, Appendix Table A6 restricts the sample for object novelty and object creativity specifications to publications with abstracts of 50 to 300 words, mitigating concerns that phrase-based novelty measures may be unreliable for extremely short or long abstracts. The estimated AMEs remain positive and highly significant across all specifications, and the mode-level ordering is preserved, confirming that the object novelty results are not driven by outliers in abstract length.

Fifth, Appendix Table A7 excludes publications in the Computer Sciences field, where AI adoption is most concentrated and where the boundary between AI-adoption and non-AI research is least clear, to verify that the estimated effects reflect the impact of AI adoption on science more broadly rather than patterns specific to the field in which AI itself largely originates. The estimated AMEs remain positive, significant, and quantitatively close to the main estimates across all specifications (e.g., AI AME of 0.136 for recombinant 3yr creativity). However, Tool-mode effects are no longer exceeding Adaptation-mode effects for recombinant creativity while Adaptation-mode effects remain largest for object creativity. This suggests that outside the computer science domain, Adaptation-based contributions are associated with both object and recombinant creativity.

## 5 DISCUSSIONS AND CONCLUSIONS

Does artificial intelligence advance science? Our findings suggest a qualified yet broadly affirmative answer. AI-adoption research is positively and significantly associated with all four primary outcomes (i.e., recombinant novelty, object novelty, short-term citation impact, and long-term citation impact), as well as all four creativity indicators that combine novelty and impact. Yet this headline finding conceals a more consequential story about how AI advances science, which depends critically on the form AI adoption takes.

The crossover pattern across AI research modes reveals a fundamental tension in how AI generates scientific value. Tool-oriented research achieves novelty primarily through recombination: by applying existing AI models to domain tasks, it inherently bridges computational and substantive literatures, producing reference portfolios that are more atypical relative to disciplinary norms. One caveat, this structural feature may be partly (perhaps even largely) mechanical for this type of application: a paper applying a computer vision algorithm to cell biology necessarily cites both AI and biology literature, generating atypical journal co-citation combinations by construction. The low object novelty associated with Tool-mode research is the other side of this coin. When the AI model is imported wholesale from elsewhere, the conceptual vocabulary of the receiving domain need not change; the novelty lies in the connection, not in what the connection produces.

Adaptation-mode research tells a different story. Modifying or reconfiguring AI models for domain-specific problems requires researchers to build new interfaces between technical AI concepts and domain-specific phenomena, which is a process that necessarily generates new vocabulary, new constructs, and new conceptual categories. The high object novelty and citation impact associated with Adaptation-mode research suggests that genuine creative integration of AI into a scientific domain does not simply import an external tool but transforms the conceptual grammar of the field it enters. This is, in a meaningful sense, a deeper form of scientific creativity: one that changes not only what combinations are made but what can be thought and named.

Foundational AI research occupies a distinct position. The consistently positive but relatively modest effects across all outcomes likely reflect the insular dynamics of frontier AI research: contributions to core AI methodology are evaluated within a specialized community whose citation norms differ from interdisciplinary contexts, and whose conceptual innovations may take longer to propagate through citing fields. The relatively compressed effects for Foundational research are therefore not evidence of lower scientific value but of a different temporal and social structure of knowledge diffusion.

Taken together, these patterns suggest that AI's relationship to scientific creativity is not a single mechanism but a family of mechanisms, each corresponding to a different form of human-AI epistemic engagement. Whether AI acts as an instrument of cross-domain search, a catalyst for conceptual reinvention, or a generator of foundational methods depends on choices made at the level of research design and disciplinary community.

This study has limitations. Our identification of AI-related research depends on publications that explicitly reference AI methods, models, or techniques. Accordingly, we capture only AI uses embedded in scientific content and methodology. We do not observe the expanding use of AI as a general-purpose tool for writing, editing, or ideation. Such uses may affect productivity, communication, and style, but they are not visible in publication records in a way that allows systematic measurement. Thus, our results pertain to AI as a scientific instrument and epistemic technology, not as background support for scholarly writing or idea generation. Our creativity measures are necessarily indirect and capture only observable novelty and impact. While recombinant and object novelty reflect important aspects of scientific creativity, they cannot fully represent deeper conceptual shifts, paradigm changes, or tacit knowledge creation. AI's influence on these processes may emerge over longer time horizons than those observed here. Finally, because our analysis relies on published research, it reflects both knowledge production and its selection through peer review and editorial processes. The observed associations between AI and creativity therefore combine genuine epistemic effects with the preferences of evaluation systems shaping scientific visibility.

While the present study focuses on the overall relationship between AI adoption and creative outcomes, the heterogeneity documented in prior work suggests that the creative impact of AI is unlikely to be uniform across individuals. Prior research documents substantial heterogeneity in the effects of AI across workers, including differences in productivity gains (Brynjolfsson et al., 2025; Kanazawa et al., 2026; Wang et al., 2023), output quality (Dell'Acqua et al., 2023), and task allocation (Hoffmann et al., 2024). Less experienced or lower-skilled workers may benefit more from AI assistance than more experienced or higher-skilled individuals. However, in science, which requires not only specialized knowledge but also judgement in choosing scientific paths (now made more numerous by AI) and in interpreting results, researchers with experience and higher skill can potentially be very effective AI users. Possibly, AI may function similarly to earlier waves of information technology as an equalizing force (Ding et al., 2010), narrowing performance gaps by enabling broader access to knowledge and problem-solving capabilities. AI may broaden participation in idea generation by lowering capability barriers, thereby increasing diversity in knowledge recombination, but it may also differentially affect individuals depending on their prior expertise, cognitive strategies, or domain knowledge. Understanding whether AI ultimately amplifies or narrows disparities in scientific creativity therefore remains an important open question. Future research should examine how the effects of AI on creativity vary across worker characteristics, experience levels, and organizational contexts to better identify the conditions under which AI advances scientific discovery.

## Statements and Declarations

### Conflict of interests

The authors declare no competing interests.

### Acknowledgements

We are grateful for feedback received from participants at the Technology Transfer Society Conference (2025) and Innotech (2026).

### Funding

The authors acknowledge support from the University of Manchester Faculty of Humanities Research Investment Fund, UKRI Metascience Unit (UKRI2597), and Japan Society for the Promotion of Science (Kaken 24K00278).

### Author contributions

Liangping Ding: Conceptualization; Methodology; Data curation; Formal Analysis; Writing; Funding acquisition. Cornelia Lawson: Conceptualization; Methodology; Formal analysis; Writing; Supervision; Funding acquisition; Philip Shapira: Conceptualization; Methodology; Writing; Supervision; Funding acquisition

## TABLES

**Table 1 Summary Statistics**

| | N | Mean | SD | Min | Max |
|---|---|---|---|---|---|
| Novelty Recomb. | 1127716 | 0.172 | 0.377 | 0.000 | 1.000 |
| Novelty Object | 742758 | 0.214 | 0.410 | 0.000 | 1.000 |
| Impact 3-yr | 696851 | 0.304 | 0.460 | 0.000 | 1.000 |
| Impact 10-yr | 1127716 | 0.286 | 0.452 | 0.000 | 1.000 |
| Creativity Recomb. 3yr | 696851 | 0.104 | 0.305 | 0.000 | 1.000 |
| Creativity Recomb. 10yr | 1127716 | 0.098 | 0.297 | 0.000 | 1.000 |
| Creativity Object 3yr | 532740 | 0.128 | 0.334 | 0.000 | 1.000 |
| Creativity Object 10yr | 742758 | 0.104 | 0.305 | 0.000 | 1.000 |
| AI | 1127716 | 0.619 | 0.486 | 0.000 | 1.000 |
| Adaptation | 1127716 | 0.253 | 0.435 | 0.000 | 1.000 |
| Foundational | 1127716 | 0.057 | 0.231 | 0.000 | 1.000 |
| Tool | 1127716 | 0.309 | 0.462 | 0.000 | 1.000 |
| ln(No. Phrases) | 740824 | 2.709 | 0.663 | 0.000 | 6.057 |
| ln(Reference Count) | 1127716 | 3.270 | 0.728 | 1.099 | 7.439 |
| Journal | 1127716 | 0.850 | 0.357 | 0.000 | 1.000 |
| ln(No. Authors) | 1127716 | 1.300 | 0.613 | 0.000 | 7.984 |
| ln(No. Affiliations) | 1127716 | 0.653 | 0.661 | 0.000 | 5.030 |
| ln(No. Countries) | 1127716 | 0.196 | 0.370 | 0.000 | 3.555 |

Note: The estimation sample used in this table is the recombinant novelty estimation sample which is the same as the 10-year citation impact estimation sample. Sample sizes vary across specifications due to missing values in the dependent variable: the three-year citation impact and recombinant creativity specifications are restricted to publications before 2022 (N = 696,851), while the object novelty and object creativity specifications are limited to publications with available phrase data from Arts et al. (2025).

**Table 2. Mean Differences between AI and non-AI Publications**

| | Non-AI | AI | Adaptation | Foundational | Tool |
|---|---|---|---|---|---|
| Novelty Recomb. | 0.096 | 0.218 [a] | 0.170 [a] | 0.152 [ai] | 0.270 [ahx] |
| Novelty Object | 0.224 | 0.206 [b] | 0.234 [a] | 0.234 [a] | 0.179 [biy] |
| Impact 3-yr | 0.283 | 0.322 [a] | 0.343 [a] | 0.363 [ah] | 0.297 [aiy] |
| Impact 10-yr | 0.261 | 0.302 [a] | 0.308 [a] | 0.333 [ah] | 0.291 [aiy] |
| Creativity Recomb. 3yr | 0.070 | 0.135 [a] | 0.106 [a] | 0.098 [ai] | 0.165 [ahx] |
| Creativity Recomb. 10yr | 0.062 | 0.120 [a] | 0.095 [a] | 0.083 [ai] | 0.148 [ahx] |
| Creativity Object 3yr | 0.114 | 0.142 [a] | 0.171 [a] | 0.161 [ai] | 0.118 [aiy] |
| Creativity Object 10yr | 0.100 | 0.107 [a] | 0.125 [a] | 0.127 [a] | 0.090 [biy] |

Note: The table reports mean values for non-AI and AI publications. Differences are calculated using two-sample t-tests. Statistical significance is indicated as **a** = significantly larger than non-AI ($p<0.01$); **b**=significantly smaller than non-AI ($p<0.01$); **h** = significantly larger than Adaptation ($p<0.01$); **i**= significantly smaller than Adaptation ($p<0.01$); **x**= significantly larger than Foundational ($p<0.01$); **y**= significantly smaller than Foundational ($p<0.01$)

**Table 3 Main Results of AI Adoption on Novelty and Impact**

| | (1) | (2) | (3) | (4) | (5) | (6) | (7) | (8) |
|---|---|---|---|---|---|---|---|---|
| | Novelty Recomb. | | Novelty Object | | Impact 3-yr | | Impact 10-yr | |
| AI | 0.161*** | | 0.0501*** | | 0.0889*** | | 0.0963*** | |
| | (0.000747) | | (0.00103) | | (0.00115) | | (0.000896) | |
| Adaptation | | 0.120*** | | 0.102*** | | 0.120*** | | 0.116*** |
| | | (0.00102) | | (0.00146) | | (0.00157) | | (0.00117) |
| Foundational | | 0.103*** | | 0.0732*** | | 0.0843*** | | 0.0864*** |
| | | (0.00174) | | (0.00233) | | (0.00232) | | (0.00185) |
| Tool | | 0.191*** | | 0.0225*** | | 0.0747*** | | 0.0872*** |
| | | (0.000887) | | (0.00111) | | (0.00127) | | (0.000983) |
| ln(Reference Count) | 0.0155*** | 0.0188*** | | | 0.209*** | 0.208*** | 0.191*** | 0.191*** |
| | (0.000521) | (0.000522) | | | (0.000754) | (0.000759) | (0.000625) | (0.000628) |
| ln(No. Authors) | 0.00816*** | 0.00604*** | 0.0181*** | 0.0192*** | 0.0930*** | 0.0930*** | 0.0762*** | 0.0762*** |
| | (0.000701) | (0.000697) | (0.000904) | (0.000907) | (0.000994) | (0.000996) | (0.000786) | (0.000788) |
| ln(No. Affiliations) | 0.00446*** | 0.00377*** | -0.00273** | -0.00130 | 0.0112*** | 0.0120*** | 0.00866*** | 0.00933*** |
| | (0.000704) | (0.000703) | (0.000887) | (0.000886) | (0.000994) | (0.000994) | (0.000784) | (0.000784) |
| ln(No. Countries) | -0.0131*** | -0.0117*** | 0.00627*** | 0.00475*** | 0.0564*** | 0.0559*** | 0.0602*** | 0.0600*** |
| | (0.00112) | (0.00112) | (0.00141) | (0.00140) | (0.00157) | (0.00157) | (0.00121) | (0.00121) |
| Journal | 0.0353*** | 0.0287*** | -0.0373*** | -0.0365*** | 0.0866*** | 0.0902*** | 0.0979*** | 0.0997*** |
| | (0.000986) | (0.00101) | (0.00375) | (0.00374) | (0.00144) | (0.00145) | (0.00113) | (0.00113) |
| ln(No. Phrases) | | | 0.154*** | 0.154*** | | | | |
| | | | (0.000949) | (0.000954) | | | | |
| Observations | 1127716 | 1127716 | 740824 | 740824 | 696851 | 696851 | 1127716 | 1127716 |
| Pseudo R-squared | 0.061 | 0.065 | 0.118 | 0.122 | 0.178 | 0.179 | 0.155 | 0.155 |

Note: Average marginal effects (AME) are reported. Robust delta-method SEs in parentheses. All models include publication year, OECD field fixed effects. Odd cols: binary AI indicator. Even cols: AI mode dummies (ref: non-AI). Observations with zero phrases are excluded from the object novelty analyses because their novelty scores are mechanically set to zero by construction. In these cases, however, the underlying novelty cannot be meaningfully assessed.

* $p < 0.05$, ** $p < 0.01$, *** $p < 0.001$

**Table 4 Effect of AI on Creativity (PCRE, Top 10% Dummy)**

| | (1) | (2) | (3) | (4) | (5) | (6) | (7) | (8) |
|---|---|---|---|---|---|---|---|---|
| | Creativity Recomb. 3yr | | Creativity Recomb. 10yr | | Creativity Object 3yr | | Creativity Object 10yr | |
| AI | $0.116^{***}$ | | $0.103^{***}$ | | $0.0557^{***}$ | | $0.0488^{***}$ | |
| | (0.000812) | | (0.000574) | | (0.00102) | | (0.000773) | |
| Adaptation | | $0.103^{***}$ | | $0.0946^{***}$ | | $0.101^{***}$ | | $0.0854^{***}$ |
| | | (0.00124) | | (0.000855) | | (0.00162) | | (0.00120) |
| Foundational | | $0.0842^{***}$ | | $0.0694^{***}$ | | $0.0723^{***}$ | | $0.0603^{***}$ |
| | | (0.00192) | | (0.00141) | | (0.00239) | | (0.00184) |
| Tool | | $0.126^{***}$ | | $0.112^{***}$ | | $0.0334^{***}$ | | $0.0306^{***}$ |
| | | (0.000956) | | (0.000677) | | (0.00110) | | (0.000827) |
| ln(Reference Count) | $0.0793^{***}$ | $0.0803^{***}$ | $0.0705^{***}$ | $0.0713^{***}$ | $0.0739^{***}$ | $0.0726^{***}$ | $0.0586^{***}$ | $0.0579^{***}$ |
| | (0.000589) | (0.000591) | (0.000464) | (0.000464) | (0.000743) | (0.000747) | (0.000585) | (0.000589) |
| ln(No. Authors) | $0.0477^{***}$ | $0.0465^{***}$ | $0.0368^{***}$ | $0.0355^{***}$ | $0.0399^{***}$ | $0.0410^{***}$ | $0.0310^{***}$ | $0.0320^{***}$ |
| | (0.000723) | (0.000722) | (0.000554) | (0.000553) | (0.000897) | (0.000902) | (0.000697) | (0.000703) |
| ln(No. Affiliations) | $0.00557^{***}$ | $0.00545^{***}$ | $0.00283^{***}$ | $0.00282^{***}$ | -0.000551 | 0.000648 | $-0.00151^{*}$ | -0.000441 |
| | (0.000682) | (0.000683) | (0.000528) | (0.000529) | (0.000859) | (0.000857) | (0.000664) | (0.000663) |
| ln(No. Countries) | $0.00357^{***}$ | $0.00425^{***}$ | $0.0112^{***}$ | $0.0117^{***}$ | $0.0187^{***}$ | $0.0174^{***}$ | $0.0153^{***}$ | $0.0143^{***}$ |
| | (0.00105) | (0.00105) | (0.000795) | (0.000794) | (0.00133) | (0.00133) | (0.00101) | (0.00101) |
| Journal | $0.0555^{***}$ | $0.0536^{***}$ | $0.0597^{***}$ | $0.0581^{***}$ | $-0.0405^{***}$ | $-0.0368^{***}$ | $-0.0279^{***}$ | $-0.0255^{***}$ |
| | (0.000889) | (0.000904) | (0.000660) | (0.000672) | (0.00407) | (0.00402) | (0.00320) | (0.00317) |
| ln(No. Phrases) | | | | | $0.0994^{***}$ | $0.1000^{***}$ | $0.0832^{***}$ | $0.0836^{***}$ |
| | | | | | (0.000989) | (0.000998) | (0.000784) | (0.000791) |
| Observations | 696851 | 696851 | 1127716 | 1127716 | 531295 | 531295 | 740824 | 740824 |
| Pseudo R-squared | 0.147 | 0.148 | 0.140 | 0.141 | 0.116 | 0.120 | 0.135 | 0.140 |

Note: Average marginal effects (AME) are reported and all Dependent variables are PCRE dummy variables. Robust delta-method SEs in parentheses. All models include publication year, OECD field fixed effects. Odd cols: binary AI indicator. Even cols: AI mode dummies (ref: non-AI). PCRE = percentile rank creativity combining novelty and citation impact. Observations with zero phrases are excluded from the object-based creativity analyses because their creativity scores are mechanically set to zero by construction. In these cases, however, the underlying creativity cannot be meaningfully assessed.

$^{*}$ $p < 0.05$, $^{**}$ $p < 0.01$, $^{***}$ $p < 0.001$

**Figure 1 Dataset Construction**

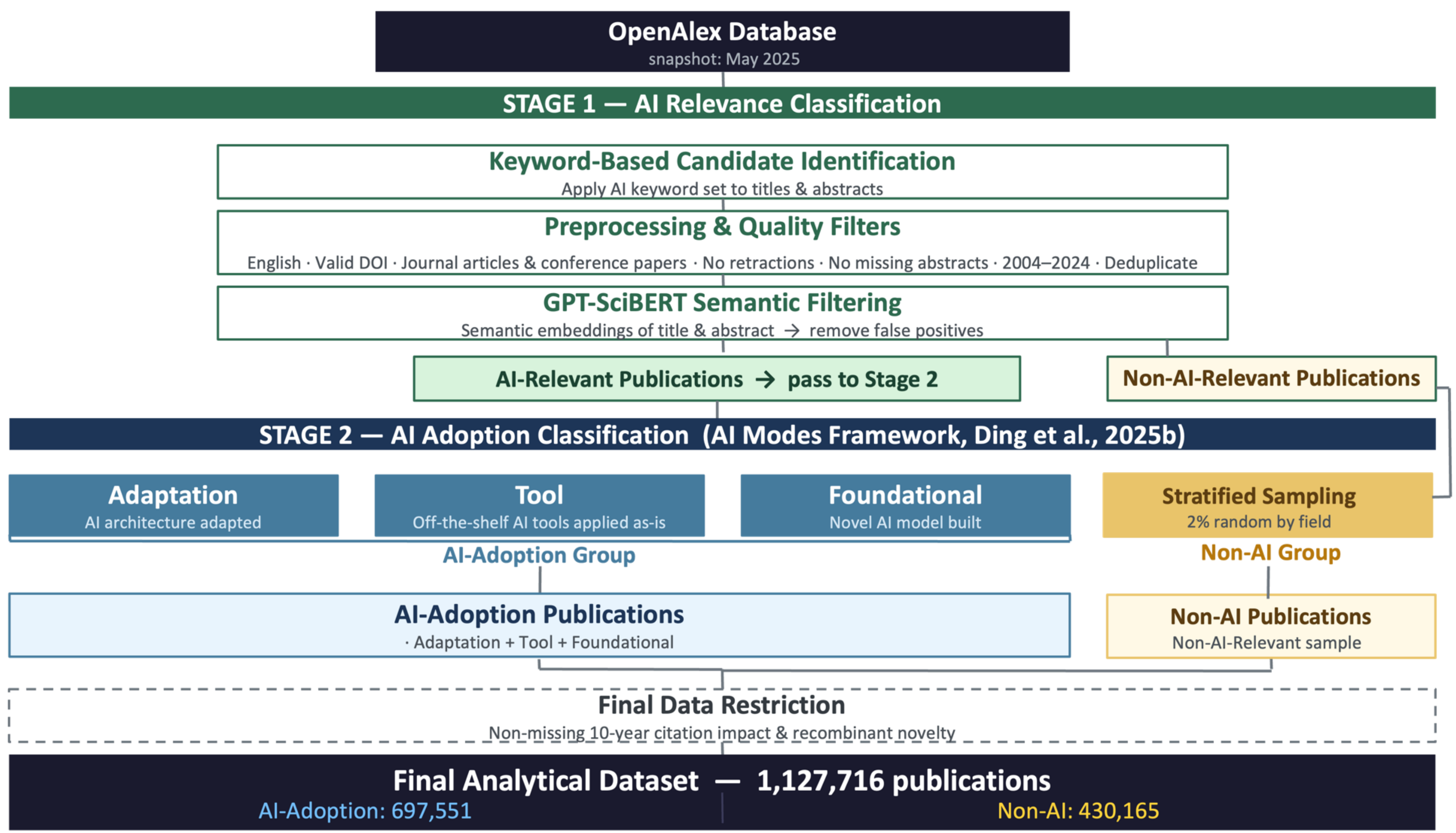

**Figure 2 Predictive Margins of Novelty and Impact Across AI modes**

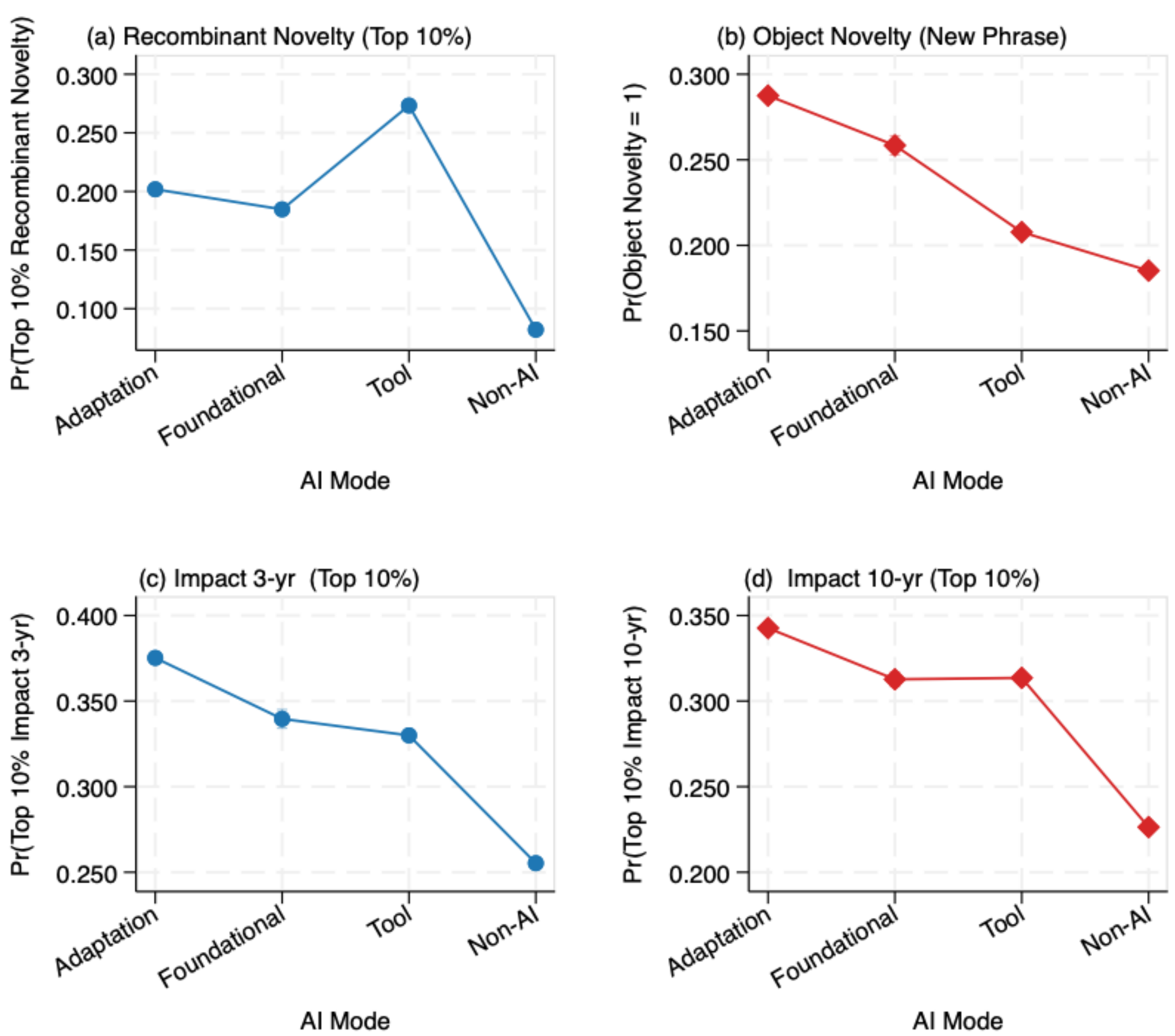

Predicted probabilities from probit with 99% CIs. Controls at means. Ref: Non-AI.

**Figure 3 Predictive Margins of Creativity Across AI modes**

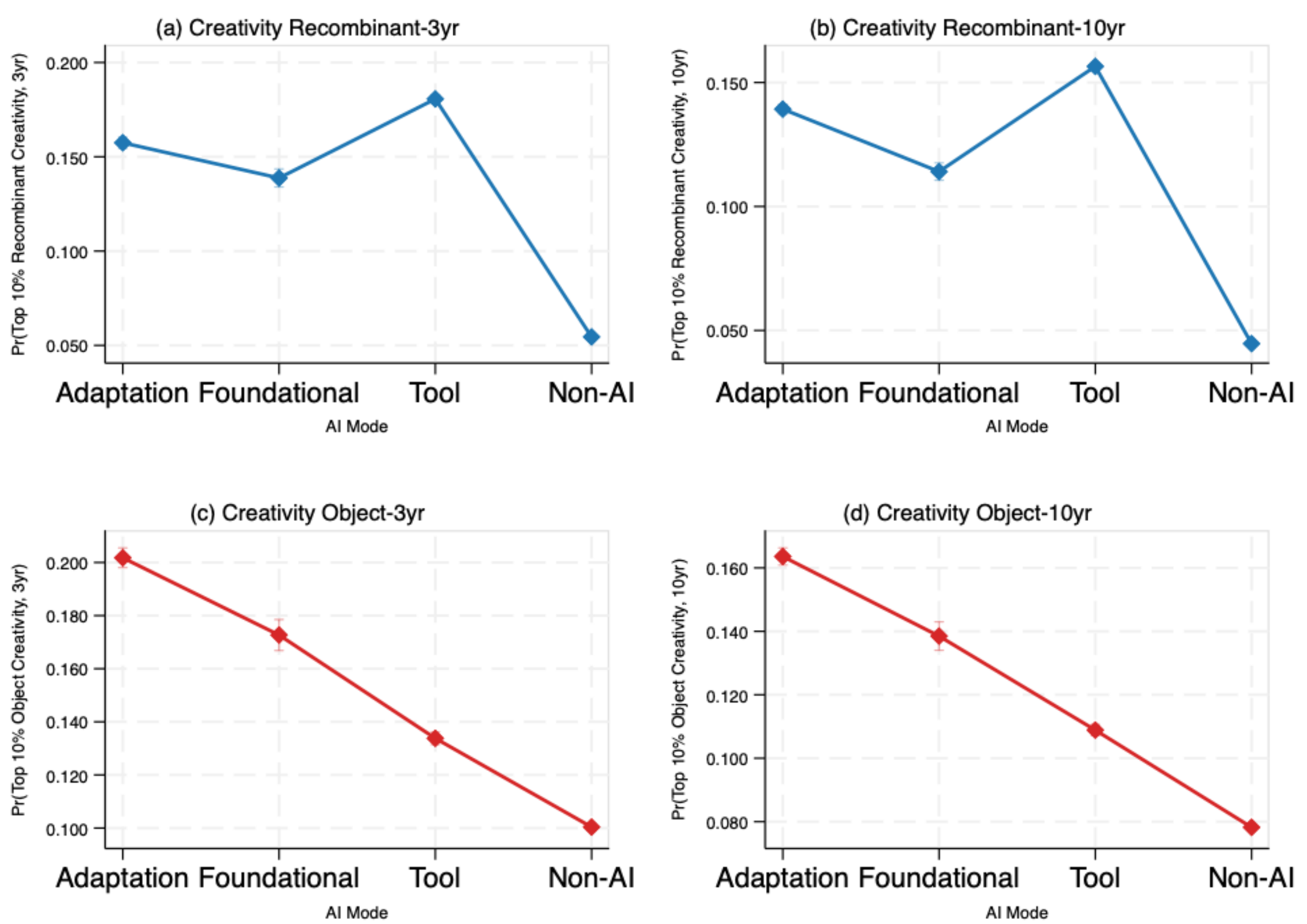

Predicted probabilities from probit with 99% CIs. Controls at means. Ref: Non-AI.

## APPENDICES

### Table A1 Variable Definitions

| Variable | Definition |
|---|---|
| ***Panel A: Outcome Variables*** | |
| Novelty Recomb. | Binary indicator equal to 1 if the publication ranks in the top 10% of recombinant novelty within its OECD field. Novelty is measured by the atypicality of publication outlets in the reference list (Lee et al., 2015). |
| Novelty Object | Binary variable equal to 1 if the paper introduces at least one new noun phrase in its title or abstract that is subsequently reused in at least one other paper, following Arts et al. (2025). |
| Impact 3-yr | Binary indicator equal to 1 if the publication ranks in the top 10% of forward citation counts within its OECD field and publication year, based on citations received within three years of publication. |
| Impact 10-yr | Binary indicator equal to 1 if the publication ranks in the top 10% of forward citation counts within its OECD field and publication year, based on citations received within ten years of publication. |
| Creativity Recomb. 3yr | Binary indicator equal to 1 if the publication ranks in the top 10% of the PCRE score combining recombinant novelty and three-year citation impact percentile ranks (each ranked 1–100, averaged, scaled 0–1). |
| Creativity Recomb. 10yr | Binary indicator equal to 1 if the publication ranks in the top 10% of the PCRE score combining recombinant novelty and ten-year citation impact percentile ranks. (each ranked 1–100, averaged, scaled 0–1). |
| Creativity Object 3yr | Binary indicator equal to 1 if the publication ranks in the top 10% of the PCRE score combining object novelty and three-year citation impact percentile ranks. (each ranked 1–100, averaged, scaled 0–1). |
| Creativity Object 10yr | Binary indicator equal to 1 if the publication ranks in the top 10% of the PCRE score combining object novelty and ten-year citation impact percentile ranks. (each ranked 1–100, averaged, scaled 0–1). |
| ***Panel B: Explanatory Variables*** | |
| AI | Binary indicator equal to 1 if the publication adopts AI methods. |
| Adaptation | Binary indicator equal to 1 if the AI-adopting publication is classified as Adaptation mode. |
| Foundational | Binary indicator equal to 1 if the AI-adopting publication is classified as Foundational mode. |
| Tool | Binary indicator equal to 1 if the AI-adopting publication is classified as Tool mode. |
| ***Panel C: Control Variables*** | |
| ln(Reference Count) | Natural logarithm of 1 plus the number of reference count. |
| ln(No. Authors) | Natural logarithm of the number of authors. |
| ln(No. Affiliations) | Natural logarithm of the number of distinct author affiliations. |
| ln(No. Countries) | Natural logarithm of the number of distinct author countries. |
| ln(No. Phrases) | Natural logarithm of 1 plus the number of noun phrases in the title and abstract of the publication. |
| Journal | Binary indicator equal to 1 if the publication type is a journal article, and 0 if conference proceeding. |

**Table A2 Pairwise Correlations**

| Variables | (1) | (2) | (3) | (4) | (5) | (6) | (7) | (8) | (9) | (10) |
|---|---|---|---|---|---|---|---|---|---|---|
| (1) AI | 1.000 | | | | | | | | | |
| (2) Adaptation | 0.457 | 1.000 | | | | | | | | |
| (3) Foundational | 0.192 | -0.143 | 1.000 | | | | | | | |
| (4) Tool | 0.525 | -0.389 | -0.164 | 1.000 | | | | | | |
| (5) ln(Reference Count) | 0.028 | 0.014 | 0.047 | -0.008 | 1.000 | | | | | |
| (6) ln(No. Authors) | 0.021 | -0.009 | -0.052 | 0.057 | 0.237 | 1.000 | | | | |
| (7) ln(No. Affiliations) | -0.037 | -0.063 | -0.005 | 0.024 | 0.245 | 0.497 | 1.000 | | | |
| (8) ln(No. Countries) | 0.019 | -0.004 | 0.027 | 0.010 | 0.183 | 0.283 | 0.560 | 1.000 | | |
| (9) ln(No. Phrases) | 0.023 | 0.012 | -0.062 | 0.045 | 0.223 | 0.258 | 0.135 | 0.064 | 1.000 | |
| (10) Journal | -0.229 | -0.191 | -0.103 | -0.010 | 0.263 | 0.046 | 0.122 | 0.051 | 0.053 | 1.000 |

Note: Pairwise Pearson correlations. Sample: recombinant novelty/10 year citation impact estimation. All of the variance inflation factors (VIFs) remain below the conventional threshold of 10, supporting the inclusion of these variables jointly in the regression models.

## Table A3: Robustness — Factor Analysis Creativity (Top 10%)

| | Creativity Recomb. 3yr | Creativity Recomb. 10yr | Creativity Object 3yr | Creativity Object 10yr |
|---|---|---|---|---|
| | (1) | (3) | (5) | (7) |
| AI | 0.114*** | 0.0987*** | 0.0522*** | 0.0458*** |
| | (0.000841) | (0.000589) | (0.00102) | (0.000769) |
| | (2) | (4) | (6) | (8) |
| Adaptation | 0.0974*** | 0.0844*** | 0.100*** | 0.0811*** |
| | (0.00125) | (0.000847) | (0.00163) | (0.00119) |
| Foundational | 0.0802*** | 0.0614*** | 0.0726*** | 0.0583*** |
| | (0.00190) | (0.00136) | (0.00241) | (0.00182) |
| Tool | 0.127*** | 0.111*** | 0.0284*** | 0.0280*** |
| | (0.000994) | (0.000698) | (0.00109) | (0.000822) |
| Observations | 696851 | 1127716 | 531295 | 740824 |

Note: Average marginal effects (AME) reported. Robust delta-method SEs in parentheses. All models include publication year, OECD field fixed effects. All controls included. DV: top-10% dummy based on factor analysis combining novelty and citation impact. Odd cols: binary AI indicator. Even cols: AI mode dummies (ref: non-AI). $^{*}$ $p < 0.05$, $^{**}$ $p < 0.01$, $^{***}$ $p < 0.001$

**Table A4: Robustness — Continuous Creativity (PCRE)**

| | Creativity Recomb. 3yr | Creativity Recomb. 10yr | Creativity Object 3yr | Creativity Object 10yr |
|---|---|---|---|---|
| | (1) | (3) | (5) | (7) |
| AI | 0.0985*** | 0.106*** | 0.0516*** | 0.0548*** |
| | (0.000504) | (0.000408) | (0.000708) | (0.000562) |
| | (2) | (4) | (6) | (8) |
| Adaptation | 0.0766*** | 0.0847*** | 0.0873*** | 0.0818*** |
| | (0.000686) | (0.000529) | (0.00101) | (0.000753) |
| Foundational | 0.0703*** | 0.0709*** | 0.0554*** | 0.0519*** |
| | (0.00105) | (0.000854) | (0.00158) | (0.00125) |
| Tool | 0.112*** | 0.120*** | 0.0353*** | 0.0423*** |
| | (0.000552) | (0.000442) | (0.000772) | (0.000606) |
| Observations | 696851 | 1127716 | 531295 | 740824 |

Note: OLS coefficients. Robust delta-method SEs in parentheses. All models include publication year, OECD field fixed effects. DV: percentile rank creativity score (continuous, 0-1). Odd cols: binary AI indicator. Even cols: AI mode dummies (ref: non-AI).
$^{*}$ $p < 0.05$, $^{**}$ $p < 0.01$, $^{***}$ $p < 0.001$

### Table A5: Robustness — Excluding US and China Authors

| | Creativity Recomb. 3yr | Creativity Recomb. 10yr | Creativity Object 3yr | Creativity Object 10yr |
|---|---|---|---|---|
| | (1) | (3) | (5) | (7) |
| AI | 0.111*** | 0.104*** | 0.0481*** | 0.0434*** |
| | (0.00108) | (0.000781) | (0.00132) | (0.00102) |
| | (2) | (4) | (6) | (8) |
| Adaptation | 0.104*** | 0.101*** | 0.0909*** | 0.0766*** |
| | (0.00181) | (0.00127) | (0.00237) | (0.00178) |
| Foundational | 0.0832*** | 0.0718*** | 0.0587*** | 0.0473*** |
| | (0.00280) | (0.00212) | (0.00338) | (0.00263) |
| Tool | 0.117*** | 0.109*** | 0.0334*** | 0.0320*** |
| | (0.00122) | (0.000880) | (0.00139) | (0.00108) |
| Observations | 386166 | 607216 | 286744 | 387011 |

Note: Average marginal effects (AME) from probit. Robust delta-method SEs in parentheses. All models include publication year, OECD field fixed effects. Sample excludes papers with at least one US or Chinese co-author. Odd cols: binary AI indicator. Even cols: AI mode dummies (ref: non-AI).
$^{*}$ $p < 0.05$, $^{**}$ $p < 0.01$, $^{***}$ $p < 0.001$

**Table A6: Robustness — Abstract Word Count 50-300 Words**

| | Novelty Object | Creativity Object 3yr | Creativity Object 10yr |
|---|---|---|---|
| | (1) | (3) | (5) |
| AI | 0.0700$^{***}$ | 0.0903$^{***}$ | 0.0703$^{***}$ |
| | (0.00183) | (0.00214) | (0.00147) |
| | (2) | (4) | (6) |
| Adaptation | 0.135$^{***}$ | 0.167$^{***}$ | 0.125$^{***}$ |
| | (0.00257) | (0.00347) | (0.00230) |
| Foundational | 0.108$^{***}$ | 0.117$^{***}$ | 0.0872$^{***}$ |
| | (0.00471) | (0.00575) | (0.00402) |
| Tool | 0.0429$^{***}$ | 0.0628$^{***}$ | 0.0490$^{***}$ |
| | (0.00191) | (0.00224) | (0.00153) |
| Observations | 263829 | 168011 | 263829 |

Note: Average marginal effects (AME) from probit. Robust delta-method SEs in parentheses. All models include publication year, OECD field fixed effects. Sample: object novelty papers with abstracts of 50-300 words. Odd cols: binary AI indicator. Even cols: AI mode dummies (ref: non-AI).
$^{*}$ $p < 0.05$, $^{**}$ $p < 0.01$, $^{***}$ $p < 0.001$

**Table A7: Robustness — Excluding Computer Science Field**

| | Creativity Recomb. 3yr | Creativity Recomb. 10yr | Creativity Object 3yr | Creativity Object 10yr |
|---|---|---|---|---|
| | (1) | (3) | (5) | (7) |
| AI | 0.136*** | 0.121*** | 0.0595*** | 0.0522*** |
| | (0.000993) | (0.000708) | (0.00110) | (0.000841) |
| | (2) | (4) | (6) | (8) |
| Adaptation | 0.140*** | 0.126*** | 0.115*** | 0.0967*** |
| | (0.00168) | (0.00114) | (0.00198) | (0.00146) |
| Foundational | 0.115*** | 0.0955*** | 0.0819*** | 0.0673*** |
| | (0.00318) | (0.00235) | (0.00367) | (0.00284) |
| Tool | 0.136*** | 0.122*** | 0.0357*** | 0.0328*** |
| | (0.00113) | (0.000804) | (0.00119) | (0.000907) |
| Observations | 540019 | 861974 | 437900 | 600892 |

Note: Average marginal effects (AME) from probit. Robust delta-method SEs in parentheses. All models include publication year, OECD field fixed effects. Sample excludes papers in Computer Sciences (OECD field). Odd cols: binary AI indicator. Even cols: AI mode dummies (ref: non-AI).
* $p < 0.05$, ** $p < 0.01$, *** $p < 0.001$